# People's Water Data: A Scalable Student-Led Methodology for Global Water Security

Suzan Kagan [a,b], Sankar Sudhir [a], Ben Hamilton [c], Ramya Dwivedi [a], P. Amaravathy [d], Alona Maslennikov [b], Timor Lichtman [f], Sonali Seth [a], Tanmayaa Nayak [a], Ganesan Velmurugan [d,e], Hadas Mamane [b], Ram Fishman [f], Thalappil Pradeep [a]

[a] DST Unit of Nanoscience and Thematic Unit of Excellence, Department of Chemistry, IIT Madras, Chennai 600 036, India.

[b] School of Mechanical Engineering, Faculty of Engineering, Tel Aviv University, Tel-Aviv, Israel.

[c] Faculty of Environment, Science and Economy, University of Exeter, Exeter, United Kingdom.

[d] Chemomicrobiomics Laboratory, Department of Biochemistry and Microbiology, KMCH Research Foundation, Coimbatore 641 014, Tamil Nadu, India.

[e] Central Research Laboratory, KMCH Institute of Health Science and Research, Coimbatore 641 014, Tamil Nadu, India.

[f] Sustainable Development Program, Tel Aviv University, Tel-Aviv, Israel.

**Corresponding Author**
Thalappil Pradeep
Email: pradeep@iitm.ac.in
Tel.: +91-44 2257 4208; Fax: +91-44 2257 0545/0509
3International Centre for Clean Water, 2nd Floor, B-Block, IIT Madras Research Park, Kanagam Road, Taramani, Chennai 600113, India

**Abstract**

Ensuring access to safe drinking water, the most precious resource for development, remains a global challenge, particularly in regions with limited infrastructure and low public awareness. The People's Water Data (PWD) initiative addresses this gap by democratizing science and empowering non-experts, particularly students, to act as agents of change in their communities. Rather than simply reporting water quality data, PWD introduces a scalable framework that combines scientific training with community engagement to foster water safety practices at the grassroots. In this study, over 1,600 students were trained through a hybrid curriculum that integrated theoretical learning with practical fieldwork. Of these, 990 participants advanced to hands-on sampling, collectively surveying more than 9,000 household drinking water sources. Using simple, standardized tools, they delivered personalized water quality reports and recommendations to families, facilitating immediate improvements in daily practices such as filtration and safe storage. The findings emphasize how scientific capacity was built effectively and translated into action. By fostering water literacy and community trust, PWD aims to catalyze behavioral change and create localized pathways toward safer water use. This participatory approach demonstrates that empowering individuals with accessible tools and knowledge can generate reliable environmental data while transforming public engagement and governance from the bottom up. The PWD model thus provides a replicable template for student-led environmental monitoring, delivering both credible insights and lasting social impact for sustainable resource management.

**Keywords:** People's Water Data (PWD); Drinking water quality; Citizen science; Household surveys; Hybrid training; Student-led monitoring; *E. coli*; Sustainable resource management; Global water security.

# 1. Introduction

According to the World Health Organization (WHO), an estimated 1.7 billion people worldwide relied on drinking water sources contaminated with feces, in 2022.[1] Approximately one in five children lacks access to sufficient drinking water to meet the daily needs.[1,4] Per WHO guidelines, water intended for human consumption must be free from microbiological agents that cause diseases.[2] Consumption of contaminated drinking water significantly contributes to the global burden of disease, causing an estimated 1.8 to 2.5 million deaths annually among children under five.[3] Household water treatment (HWT) practices, when properly implemented and maintained are effective in eliminating or inactivating pathogenic microorganisms. Such treatments are particularly beneficial in settings where access to safely managed piped water is limited, ensuring the supply of microbiologically safe water at the point of use.[1]

Several studies have investigated water quality alongside household perceptions, behaviors, and treatment practices, highlighting their impact on public health outcomes. These efforts have provided important insights into how communities interact with and respond to drinking water issues. To capture both subjective factors (user behavior, practices, and perceptions) and objective indicators (quantitative water quality data), an integrated methodology combining household surveys with water quality analysis was developed.[5] Despite this progress, a clear gap remains in understanding the effectiveness of common household water treatment systems in urban settings, particularly regarding user behaviors, perceptions, maintenance, and the water quality at the point of use.

In Bekasi, Indonesia, a study implemented a participatory monitoring approach to assess microbial water quality in self-supply services. Households conducted biweekly *Escherichia coli* tests over six months, revealing contamination rates between 11% to 70% at the source and between 15% to 44% at the point of use. While the initiative increased water safety awareness and knowledge, nearly half of the participants' households dropped out, and increased awareness did not consistently translate into improved water safety behaviors. These findings suggest that household-led monitoring can support community-based water quality assessment; however, its effectiveness depends on sustained support for safe water treatment and storage and constant evaluation of long-term impacts.[6]

The Drinking Water Tool (DWT), developed in California, represents a targeted initiative to improve access to water quality data for domestic well users, particularly in rural and socioeconomically disadvantaged areas. Aimed at promoting environmental justice, this tool enhances transparency and informed decision-making in regions with limited monitoring infrastructure. Stakeholder engagement played a crucial role in refining the DWT, with community involvement through organizations like the Community Water Center helping build trust and facilitating acceptance. Despite its success, among the identified limitations was the tool's reliance on generalized data, as well as agencies trust and coordination in adoption the tool.[7]

In the Netherlands, The Freshness of Water citizen science project, engaged 43 participants in monitoring the microbiological stability of their household drinking water. Participants collected kitchen tap water samples and conducted home-based microbial tests, while also providing survey feedback. Notably, 35% reported adopting increased precautionary behaviors in their water safety practices, highlighting the potential of non-expert involvement not only as a means of data collection but also as an effective intervention for promoting health-conscious behaviors through direct engagement and experiential learning.[8]

Based on the EGUsphere (2024) initiative, a community-based water quality monitoring model piloted in Southern African countries. Community members were trained to measure basic water quality parameters such as pH, turbidity, conductivity, and dissolved oxygen using low-cost test kits. The participants collected and reported data with adequate reliability, particularly when supported by local authorities. Participants were able to collect and report data with adequate reliability, particularly when supported by local authorities. The initiative fostered increased water safety awareness, community ownership of water resources, and improved collaboration with government agencies. Despite certain challenges related to measurement accuracy and the need for sustained training, the study demonstrated that community-based monitoring can serve as a viable and socially empowering tool for environmental surveillance.[9]

While previous studies have provided important insights, none have produced a comprehensive tool that simultaneously integrates objective water quality parameters with household-level behavioral and perceptual data. The People's Water Data (PWD) initiative presented in this study represents a step toward filling this gap by combining large-scale community engagement spread over a large geographical area with systematic water quality monitoring. The PWD initiative, conducted across India, Uganda, and Israel, trained 1,690 students in the theory phase. Of these, 990 participants were involved in the field phase, collectively evaluating the quality of 9,067 water sources, primarily at the household level.

In many low- and middle-income settings, the absence of reliable household-level water quality data remains a major constraint to achieving sustainable water resource management. People's Water Data (PWD) addresses this gap through an integrated, student-led monitoring model that combines education, low-cost testing, and digital data sharing. By training university students to collect, analyze, and upload drinking water results directly from households, PWD generates spatially distributed, open-access datasets that enhance both scientific understanding and local decision-making capacity. This approach transforms data collection into a participatory process that strengthens community engagement while maintaining scientific rigor. Early implementation across India, Israel, and Uganda has revealed microbial and chemical contamination patterns and informed local management responses. Beyond data generation, the initiative has advanced water literacy, encouraged behavioral change in water handling, and built capacity for long-term monitoring.

The People's Water Data (PWD) initiative reflects the core values and mission of *ACS Sustainable Resource Management* by integrating science, education, and community engagement to promote responsible use of water resources. It demonstrates how sustainability extends beyond technology through population development, behavioral change, and equitable access to knowledge. The initiative fosters both environmental responsibility and social empowerment by training students to collect, analyze, and interpret water quality data at the household level. PWD reflects the principles of sustainable resource management through its focus on low-cost, data-driven solutions that build local capacity and resilience. Its participatory framework connects scientific rigor with social inclusion, addressing not only water quality but also the systems of behavior and governance that sustain it. In doing so, the initiative advances the journal's vision of transforming resource management into an inclusive, circular, and knowledge-based practice that ensures long-term water security for all.

The methodology includes three components: (1) a comprehensive survey, accessible via mobile phones in both online and offline modes, to capture water-related habits, perceptions, and selected behavioral and physical water quality parameters; (2) on-site water sampling and measurements of key water quality indicators, supplemented by laboratory analysis, and (3) integration of all collected data into a unified platform for systematic analysis. The PWD initiate aims to provide insights into the current state of drinking water quality, particularly in low-middle-income households, and to identify strategies for improving water safety through targeted education, effective maintenance, and strengthened community engagement.

## 2. Materials and methods

As illustrated in **Figure 1**, the People's Water Data (PWD) initiative employs a structured, multi-phased methodology to ensure both scientific rigor and community engagement. Students begin with Phase 1 (1a. Theoretical training), completing 40 academic hours of instruction delivered by expert faculty through the NPTEL platform. Then, in Phase 2: Pre-field preparation, students are trained in standard operating procedures (SOPs), 2a. install the EpiCollect5 survey app, prepare field kits, and 2b. design randomized, geographically diverse sampling strategies. Next, during the Phase 3: Field phase, students 3a. conduct household surveys, 3b. perform on-site testing of water quality, and collect sterile samples for laboratory analysis. Later, in Phase 4: Laboratory analysis, the process begins with 4a. instrument calibration, 4b. parameter measurements, 4c. first data upload, 4d. biological testing after incubation, and 4e. a second data upload. Finally, in Phase 5: Integration with PWD, 5a. data cleaning and quality control precede 5b. the final platform upload. This structured pipeline enables non-experts to generate reliable and reproducible data, supporting large-scale water quality monitoring and providing evidence for policy and management applications.

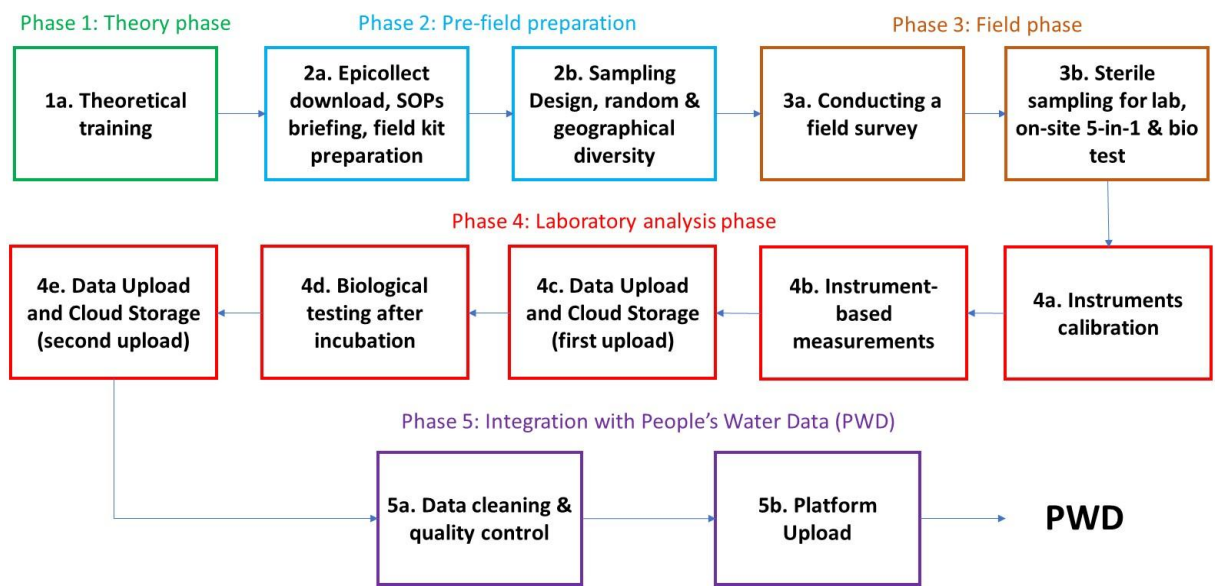

**Figure 1.** Workflow of the People's Water Data (PWD) initiative showing sequential phases from training to field sampling, laboratory analysis, and integration into the open-access platform.

## 2.1. Theoretical Training and Pre-Field Preparation

Before entering the field, all participating students completed theoretical training as part of a structured hybrid training course titled "Water Quality - An Approach to People's Water Data" (IIT Madras Course ID 5011). The course combines theoretical lectures, laboratory demonstrations, and field assignments to equip learners with both fundamental and applied skills in water quality monitoring.

It covers core modules that integrate principles of water science, sustainability, and behavioral change. Core scientific modules include Water Quality Concepts, Water Quality Parameters, and Water Quality Measurement, which introduce pathways of contamination, physicochemical and biological parameters, analytical methods, and impact fieldwork. Complementary modules expanded the learning to include Water Quality Survey Design, Water Microbiome for Sustainability, and Measurement, Behavior, and Impact, fostering interdisciplinary understanding and local relevance. The detailed syllabus, including learning outcomes, key references, and faculty contributions, is provided in **Table S1** in the Supporting Information.

The survey tool was designed to systematically collect data on drinking water quality and household water management practices, combining field measurements with information provided by respondents. To ensure data standardization and traceability, each sampled water source was assigned a unique ID, composed of the student's initials and followed by the full sampling date and time (e.g., AK202506271435). The ID links each sample to its specific location, water source, and sampling time, enabling precise sample tracking throughout field testing, laboratory analysis, and data processing. This approach supports accurate comparison across different water sources while minimizing errors and data loss. Further details on the survey tool structure are provided in **Table S2** in the Supporting Information.

### 2.1.1 Initial Assessment and Training

The training and preparation phase followed a structured, three-stage approach:

1) <u>Initial assessment and protocol adjustment</u>: the process began with a participatory assessment, including field visits to target areas, a pilot survey of 262, and ongoing protocol adjustment. This stage focused on mapping household and community drinking water sources, such as natural water bodies, wells, and public taps. This preliminary testing was used to refine the survey and field procedures. In parallel, an internal user-experience survey was conducted among research team members to evaluate the clarity, usability, and accessibility of the digital tool. Feedback from this internal assessment informed continuous refinements to both the survey structure and the field procedures.

2) <u>Local Engagement and Training</u>: In the second stage, emphasis was placed on engaging and training local student participants. Classroom and laboratory sessions covered both theoretical and practical aspects of water sampling and testing. The survey, detailed in **Table S2** in the Supporting Information, was translated into the local language to reflect regional cultural and environmental contexts.

3) <u>Community Household Survey and Water Testing</u>: In the final stage, random households were surveyed through 20-minutes visits. Research team members interviewed the household member responsible for water provision and conducted on-site water quality tests. Water samples were also collected for extended laboratory analysis.

Training and continuous feedback were prioritized to ensure data reliability. Non-expert accuracy improves with feedback and learning opportunities.[10] Accordingly, the PWD initiative incorporated field-based training and expert cross-verification to support skill development and improve the validity of student-led data.

**2.1.2 Digital Tool Development**

A custom digital survey was deployed using EpiCollect5 (https://www.epicollect.org), a mobile-based platform that allowed for real-time and offline data entry. This approach facilitated uninterrupted data collection while minimizing disruptions in areas with limited connectivity. The tool recorded household demographic information, field-based water quality observations, GPS coordinates, and photographic documentation, enhancing the accuracy of spatial mapping and data validation. By consolidating multiple data streams, the tool enabled a comprehensive assessment of household water quality trends and practices across urban and rural areas.

Mobile-based platforms and sensor-integrated tools enhance the efficiency and accuracy of water quality monitoring in decentralized settings. A mobile application integrated with sensors for monitoring pH, chlorine, total dissolved solids (TDS), and *E. coli* detection using AquaGenX kits can facilitate real-time assessment of water safety.[5] Similarly, the People's Water Data initiative leverages Epicollect for digital survey collection, coupled with sensor-based field testing to measure key chemical and biological parameters. This integration not only facilitates real-time data entry and tracking but also improves data reliability through automated recording and spatial tagging. By combining mobile technology with on-site testing, the initiative ensures systematic, scalable, and efficient water quality assessments, even in low-resource environments. Effective visualization tools enhance the understanding of water resource distribution, quality, and management. Bouman et al., 2024 introduce the Water Flow Diagram (WFD), which utilizes Sankey diagrams to represent urban water supply, consumption, treatment, and losses, providing a clear overview of water flow dynamics. This approach aligns with the People's Water Data initiative, which employs GIS-based spatial mapping and real-time survey tools to visualize water quality trends, contamination risks, and household water usage patterns. By integrating spatial data with analytical visualization techniques, these tools facilitate better decision-making, improve public engagement, and enhance the effectiveness of crowdsourced-collected water monitoring efforts.

**2.2 Sampling Strategy and Geospatial Design**

To ensure randomized and spatially representative survey coverage, polygon-based assignment was implemented. Each student was assigned a segmented map, where each segment corresponded to a defined polygon representing a specific neighborhood. Within each polygon, multiple candidate survey sites were pre-identified to ensure uniform spatial distribution.

Students were instructed to complete surveys at ten unique locations within their assigned polygon, which was shared through Google Maps to enhance geographic spread and minimize location-based bias. Data collection was conducted in accordance with a standard operating procedure (SOP), detailed in **Text S3** in the Supporting Information.

## 2.3 Field Survey and On-Site Testing

All field surveys and household interactions were conducted following ethical approval from the Institutional Ethics Committee (IEC) of the Indian Institute of Technology Madras (Approval No. IEC/2024-02/PT/16, valid from 21 June 2024 to 20 June 2027). The study, titled "A Hybrid Course on Water Quality - An Approach to People's Water Data," complied with national ethical standards for research involving human participants, as outlined by the Indian Council of Medical Research (ICMR). The complete approval letter and supporting documentation are available in **Figure S1** in the Supporting Information.

In the field phase, participants visited randomly selected households within each assigned polygon (Section 2.2). Each visit lasted about 20 minutes and involved surveying the household member responsible for water provision. If the primary respondent was unavailable, a neighboring household was surveyed in their place. Additionally, on-site water testing was conducted using 5-in-1 dip strips to test for hardness, alkalinity, free chlorine, total chlorine and pH, as well as biological kits to test for *E. coli* and coliform detection. Water samples were also collected in sterile tubes for further laboratory analysis. Questions constituting the survey are listed in **Table S2** in the Supporting Information.

## 2.4 Laboratory Analysis and Measurements

Water samples collected in sterilized tubes were brought to the laboratory within 24 h to preserve sample integrity. Standard protocols for sample collection and preservation were followed.[12] For metal analysis, Inductively Coupled Plasma Mass Spectrometry (ICP-MS) was employed, enabling high-sensitivity detection of toxic elements such as arsenic, lead, cadmium, chromium, and uranium. This broader monitoring approach enabled us to identify hidden risks and provide safer water insights to more vulnerable communities, strengthening our impact on public health, as detailed in **Text S4** in the Supporting Information. Details on the portable sensors and test strips for fielded tests as well as detailed lists of parameters, testing methods, and standards are provided in **Table S3** and **Table S4**, respectively, in the Supporting Information.[14]

## 2.5 Cross Verification and Validation

To overcome challenges related to data accuracy and reliability when assessments are conducted by non-experts, the PWD initiative implements expert cross-verification, GPS-tracking, and real-time data monitoring. A subset of households was revisited to recollect and compare data, allowing for systematic assessment of consistency between student and expert results. The polygon-based framework ensured spatial alignment of the comparison. This process revealed agreement rates and identified discrepancies, reinforcing the importance of structured methodologies and expert oversight in enhancing the reliability of crowdsourced data. A detailed Standard Operating Procedure (SOP) outlining each step of the process is available in **Text S3** of the Supporting Information.

## 2.6 Spatial Coverage and Survey Duration Analysis

### 2.6.1 Distance and coverage analysis

The implementation of the PWD model focused on underserved semi-urban and rural areas in India, Uganda, and Israel, regions where access to safe household-sourced drinking water remains limited. Institutions were selected based on their engagement with environmental education and capacity for interdisciplinary, community-centered learning. Target areas were prioritized based on documented water quality concerns (e.g., microbial or chemical contamination) and institutions' readiness to support localized interventions. This approach ensured scientific rigor, community relevance, and scalability of the decentralized training and monitoring model used, while also fostering long-term academic and political partnerships critical for systemic change. Thus, in selected areas, such as Nallampatti, the model also served as a platform for targeted action.

Following a comprehensive village-level mapping and diagnosis of water quality, tailored solutions were introduced, allowing for real-time observation of behavioral shifts and improvements in water management. In each study area, approximately 10% of the allocated surveys were first conducted as pilot tests to capture local needs and contextual challenges. Insights from these pilots were then used to refine the full-scale survey and intervention design, ensuring that the subsequent implementation was both accurate and locally relevant. These targeted implementations serve as regional proof-of-concept sites where the social impact of interventions can be evaluated, further informing the model's scalability and effectiveness in diverse contexts.

Geographic information system (GIS) tools were integrated into the survey methodology to support spatial planning, optimize survey distribution, and enable real-time tracking of student progress. These tools facilitated precise differentiation between urban and rural areas, ensuring proportional representation across diverse geographic contexts and enhancing the spatial representativeness of data. GIS-based spatial analysis was used to quantify the distances travelled by students between survey sites, enabling the assessment of survey activity distribution and identification of inconsistencies in site coverage.

### 2.6.2 Variability in survey duration

A layered monitoring approach was implemented to improve data reliability and reduce the risk of fabricated or low-quality data. Time-based indicators were incorporated into the survey platform to evaluate data quality and monitor compliance with protocols. These included total survey duration, time intervals between consecutive surveys conducted by the same participant, and response time per question. Short completion times were flagged as potentially rushed or fabricated responses, while longer completion times suggested more thorough engagement. Offline validation tests were also conducted to estimate expected survey durations. Real-time automated alerts identified irregularities in completion times, while real-time tracking and enhanced oversight protocols were implemented to ensure consistent adherence to standardized survey procedures.

## 2.7 Socio-Health Perspectives on Household Water Security

In addition to water quality parameters testing, the PWD methodology also integrated household health and practice surveys to capture the social dimensions of water safety. These issues highlight the need to extend drinking water quality control beyond the point of distribution to the point of consumption.[16] Research indicates that improved water sources, better hygiene practices such as hand washing and sanitation, and household-level water treatment significantly reduce diarrheal diseases in developing countries.[17] However, field studies have identified household practices that contribute to contamination or the spread of diseases, such as the use of wide-mouth containers for storage, transferring water between containers,[18] and dipping handheld utensils into water instead of using a tap or pouring method.[16]

To capture these dynamics, the PWD initiative incorporated structured health surveys as detailed in **Table S5** of the Supporting Information, such as in Coimbatore, where respondents reported on waterborne illnesses, hygiene practices, and the economic burden of disease (e.g., medical costs and lost working days). To date, 430 household health-practice surveys have been completed, providing a substantial evidence base to connect user behavior with water quality outcomes. By combining these insights with objective water quality data, the initiative followed an integrated methodology that aligns with recent frameworks emphasizing both user behavior and technical validation.[18] Positioning PWD as a socially applied model that links scientific testing with lived realities in vulnerable communities.

## 2.8 Household Feedback and Recommendations for Improved Practices

The PWD initiative combined for mobile data collection (EpiCollect5) and spatial mapping (GIS) to facilitate real-time data visualization and tracking. Each participating household received a personalized water quality report aligned with WHO standards. Reports were delivered through digital links and written in simple terms to ensure understanding. When contamination was detected, tailored recommendations were provided by volunteers to encourage behavioral changes aimed at improving water safety. Households were also encouraged to provide their own feedback through an online form. This feedback loop not only informed participants but also empowered them to improve their daily water practices, making household water use safer and more reliable. Full Water Test Report attached in **Text S6** in the Supporting Information.

The PWD methodology can also be understood through three complementary layers. Planning & Design brings together theoretical training, experimental design, the development of digital survey tools, and a pilot phase to test and adapt protocols to the local context. Field & Laboratory Execution is the practical layer where students conduct household surveys, collect drinking water samples, carry out on-site testing, and perform advanced laboratory analyses, the operational "core" of the research. Monitoring & Continuous Improvement provides quality control, and feedback loops, enabling real-time adjustments and strengthening the reliability of results. Together, these layers form a circular system where learning flows back into planning and execution, ensuring both scientific rigor and adaptability across contexts.

# 3. Results and Discussion

This section presents the key findings of the People's Water Data (PWD) initiative to evaluate the performance and impact of community-driven water monitoring. Although the methodology was implemented as a linear operational sequence (**Section 2**), the results are presented narratively to highlight spatial coverage, methodological consistency, and data reliability. Conducted across India, Uganda, and Israel, the PWD initiative trained 990 student monitors in the field phase, who collectively assessed 9,067 water sources, primarily at the household level, within a two-year period.

### 3.1 Global distribution of student-conducted water quality

**Figure S2** presents the global distribution of student-led water quality assessments conducted under the PWD initiative. Each data point represents a surveyed location, illustrating the study's geospatial coverage. A significant concentration of surveys took place in India, across both urban and rural areas, where water quality monitoring is of critical importance due to the challenges associated with infrastructure and contamination risks. Additional survey sites were distributed across southern Israel and several regions in Africa, demonstrating the initiative's international reach and adaptability across diverse socioeconomic and environmental contexts. This spatial distribution highlights the feasibility of crowdsourced water quality monitoring, where non-expert contributions, when systematically validated, can provide scientifically robust insights into global drinking water conditions.

### 3.2 Training and Pre-Field Preparation

The PWD initiative employed a two-phase training approach. Delivered as a hybrid course through NPTEL 1,690 students completed the theoretical phase, of whom 990 advanced to the field phase and collectively assessed 9,067 household-level water sources. In parallel, field staff and interns were trained to support implementation. Approximately 34 academic institutions actively participated, hosting laboratory hubs across India, Uganda, and Israel to ensure standardized training and testing.

## 3.3 Sampling Strategy and Geospatial Design

**Figure 2** examines the ability of experts to return to the same locations where students previously conducted surveys and perform new assessments to evaluate the reliability of the results. This figure presents a spatial comparison between student-collected (purple circles) and expert-collected (red crosses) water quality sampling locations within the study area. The x-axis represents Easting coordinates, while the y-axis denotes Northing coordinates, providing a geospatial reference for sample distribution. The overlap and proximity of data points indicate a high degree of agreement, above 80%, between student and expert sampling locations, suggesting a strong spatial correlation in data collection efforts. However, some divergence is evident, particularly in the lower and leftmost regions, where student-collected data appear more dispersed. This visualization highlights the effectiveness of non-expert sampling in environmental monitoring while also emphasizing the need for further validation and spatial standardization in community-driven data collection methodologies.

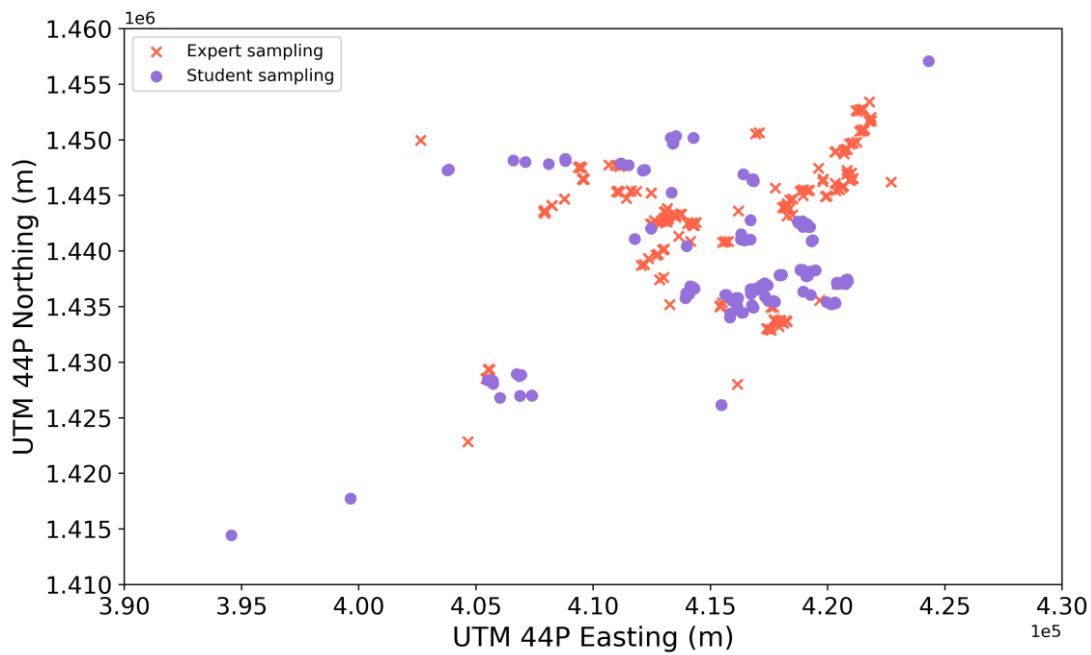

**Figure 2**. Spatial overlap between expert (red) and student (purple) water sampling locations.

To optimize geographical diversity in water quality monitoring, we implement a polygon-based random sampling approach using Sobol sequences for replicability, as illustrated in **Figure 3**, pseudo-random sampling while minimizing resampling of the same locations each year. The polygon represents the designated study area, ensuring all selected locations fall within spatial boundaries. Two allocation strategies were considered: (a) randomized student assignment, where each student selects ten location points within the polygon, and (b) K-means clustering, where locations are distributed across the entire area, ensuring students cover more distant sites. Both approaches maintain the integrity of the random allocation, as the polygon itself constrains the sampling space while allowing flexibility in surveyor movement.

For the initial trial, a region in Chennai was selected, with GPS coordinates defining the polygon boundaries. Once validated, this methodology was expanded to Coimbatore and other regions, facilitating an equitable and efficient distribution of survey locations for field researchers.

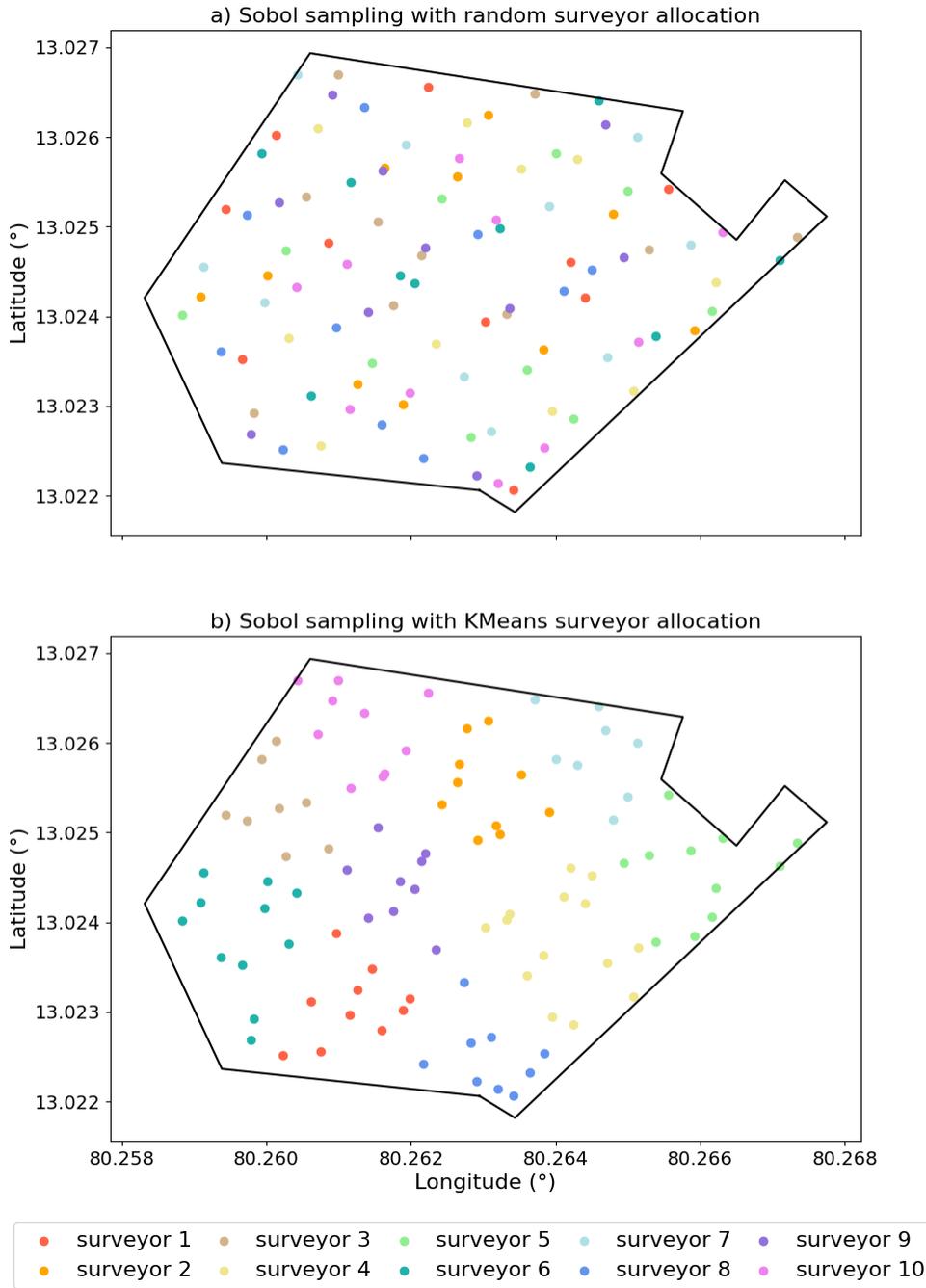

surveyor 1   surveyor 3   surveyor 5   surveyor 7   surveyor 9
surveyor 2   surveyor 4   surveyor 6   surveyor 8   surveyor 10

**Figure 3**. Spatial distribution of sampling points generated by Sobol sequence within a defined polygon: (a) random surveyor allocation and (b) K-means allocation showing improved spatial balance.

## 3.4 Spatial Coverage and Survey Duration Analysis

### 3.4.1 Distance and Coverage Analysis

**Figure 4.** presents the ten samples collected by each student, with each student represented by a different color. The chart visualizes the paths taken by students as they moved between households, providing insight into their movement patterns. Some paths are longer, indicating

that the student covered a wider area, while shorter, more concentrated paths may suggest that the data collection took place within a confined space, such as a block of buildings or a specific street. These differences may raise questions about access to water sources, walking distances, or data collection patterns.

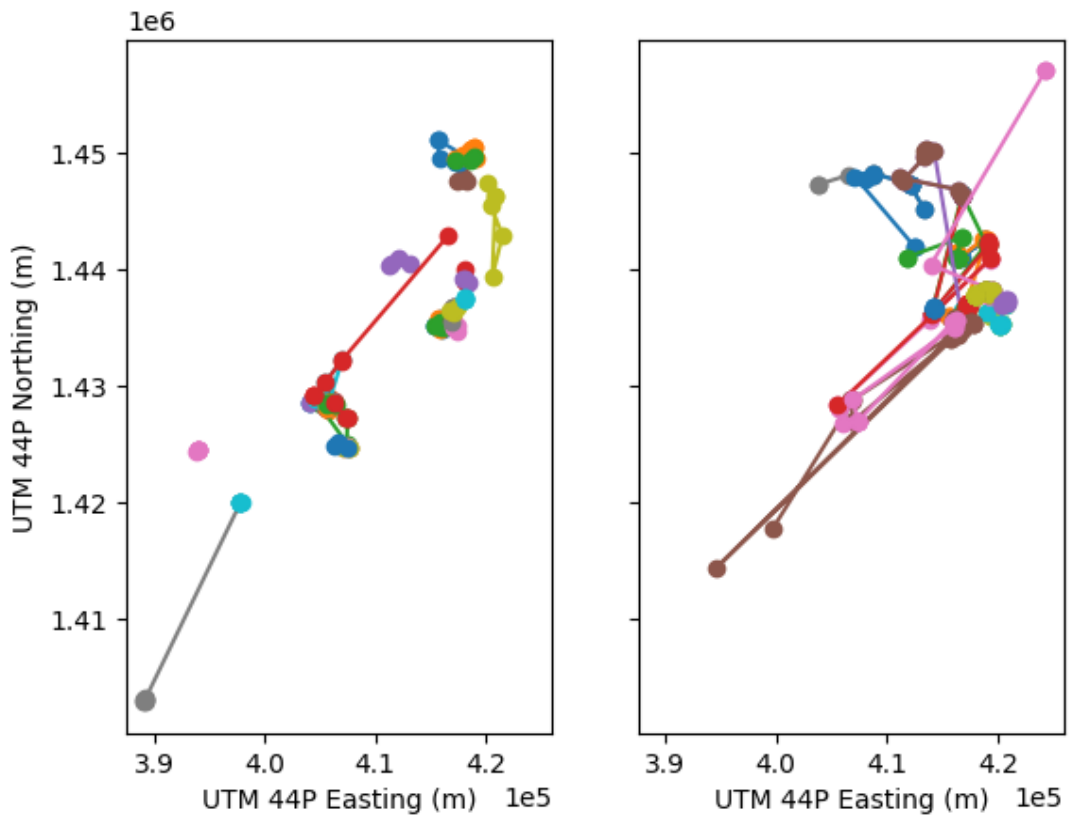

**Figure 4.** Student household water collection paths during the field campaign, with each color representing one student.

The results revealed notable variability in movement patterns, with some participants covering extensive geographic areas, while others operated within a limited spatial range. These differences highlight the need for improved consistency and enhanced methodological standardization in survey deployment. Several challenges were identified in analyzing the distances covered by students during surveys. One major issue was the uneven distribution of survey locations, with some students covering extensive areas while others remained within a limited radius. This inconsistency led to spatial biases in data collection, reducing the representativeness of the dataset. Additionally, external factors such as terrain variability, accessibility constraints, and time limitations influenced students' ability to travel between sites, further contributing to disparities in coverage. The reliance on self-reported locations also introduced potential inaccuracies, necessitating the implementation of Geographic Information System (GIS) tools to ensure precise tracking. Addressing these challenges is critical for standardizing survey coverage, improving data quality, and maintaining the integrity of the research findings.

### 3.4.2 Variability in survey duration

**Figure 5**. represents the average time it took students to answer each question. Additionally, an analysis was conducted on the time students took to complete ten surveys and the overall average survey completion time, which was found to be 22 minutes. However, some outliers were identified where students completed all ten surveys within just a few minutes. This could be due to various factors, such as working offline and uploading the data to the cloud simultaneously or the possibility that the student did not genuinely complete the surveys in real-time but rather filled them out at home, raising concerns about data reliability. **Figure 5** supports the argument that survey design influences data collection efficiency and suggests that grouping questions into logical blocks can improve response consistency.

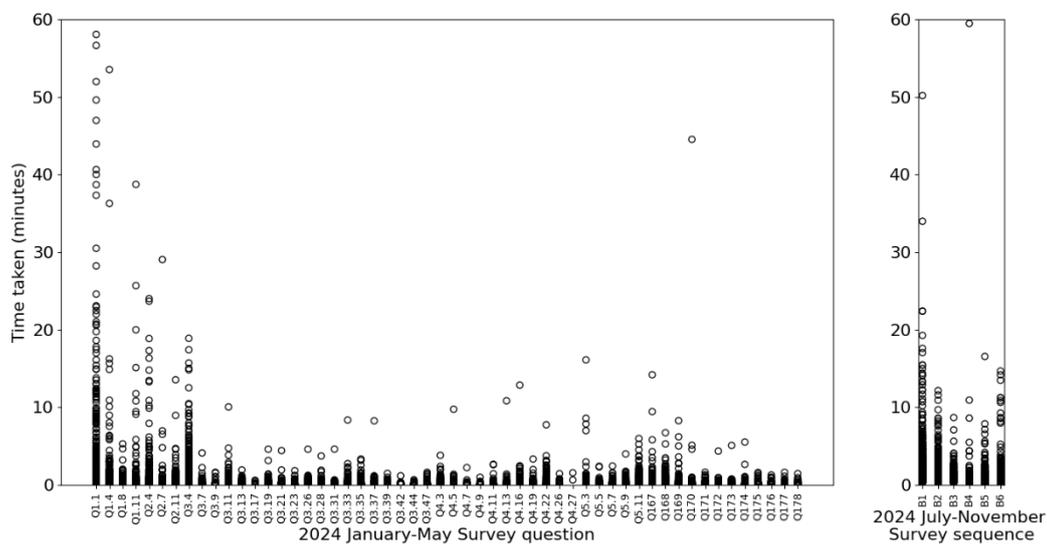

**Figure 5**. Survey completion time distribution per question and session, reflecting differences in engagement and data-entry behavior.

A comparison between the January–May and July–November survey periods reveals significant differences in time management and data consistency, driven by changes in the survey system's configuration, as shown in **Figure 6.** During the January–May semester, students were not subject to an automated cap on the number of surveys they could complete. As a result, survey participation was highly variable, with some students conducting over 20 surveys. This lack of constraint contributed to substantial variability in survey duration, including several outliers exceeding 1000 minutes, and even reaching up to 4500 minutes for individual surveys (**Figure 6.a**). Such inconsistencies raise concerns about the quality and authenticity of some entries, suggesting possible interruptions, delayed submissions, or data fabrication.

In contrast, the July–November semester introduced a system-imposed cap, requiring each student to complete exactly ten surveys. This modification led to a more standardized pattern of participation and significantly reduced variability in time taken per survey (**Figure 6.b**). Most responses fell within a reasonable timeframe (under 20 minutes), with fewer extreme outliers. The clearer distribution and lower variation suggest that the implementation of structural constraints and automation supports greater uniformity, enhances data reliability, and simplifies downstream analysis.

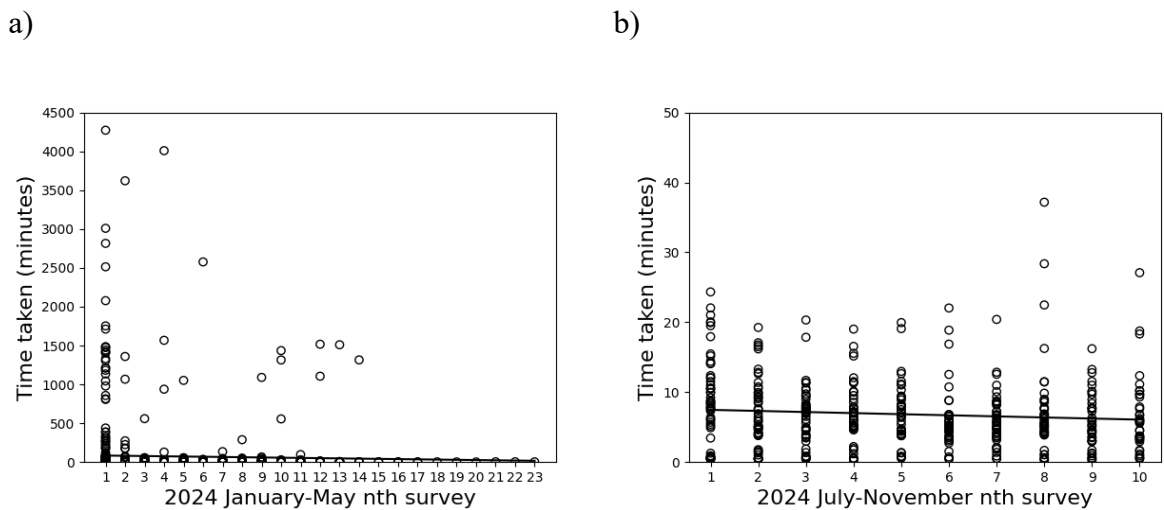

**Figure 6.** Survey duration comparison showing improved consistency after the surveys number was introduced, with limits in the July–November 2024 campaign.

Systematic validation through multiple assessments enhances data reliability by reducing individual errors and improving overall accuracy. Strobl et al.,2019 demonstrated that comparing multiple evaluations of the same data point helps filter out inconsistencies and strengthens confidence in crowdsourced data, reinforcing the "wisdom of the crowd" principle.

These findings highlight the critical role of system-level design in shaping data collection behavior. Enforcing consistent expectations and technical constraints improves both compliance and data quality in crowdsourced research frameworks.

## 3.5 Lab Analysis and Measurements

**Table S2** in the Supporting Information summarizes all water quality parameters testes in this study, including criteria from both the World Health Organization[1,4] and the Bureau of Indian Standards to ensure alignment with international and local standards.[13]

**Figure 7** displays the values of conductivity, TDS, turbidity, pH, Alkalinity, Hardness and ORP across different locations. The average electrical conductivity (EC) value was approximately 444.07 µS/cm, with a deviation of 477.57 µS/cm. The TDS mean concentrations of 186.90 mg/l. The average turbidity was around 1.08 NTU, with a range from 0 to 13.07 NTU. For pH values, the results showed no significant differences, with values ranging from 7.2 to 8.9, an average of 7.8, and a standard deviation of 0.3. ORP is a measure of the tendency of a solution to either gain or lose electrons in a chemical reaction, with values averaging -4.83 ± 60.92 mV. The average alkalinity was around 96.05 ppm, with a range from 0 to 240 ppm. The average hardness is around 144.63 ppm, with a range from 0 to 425 ppm.

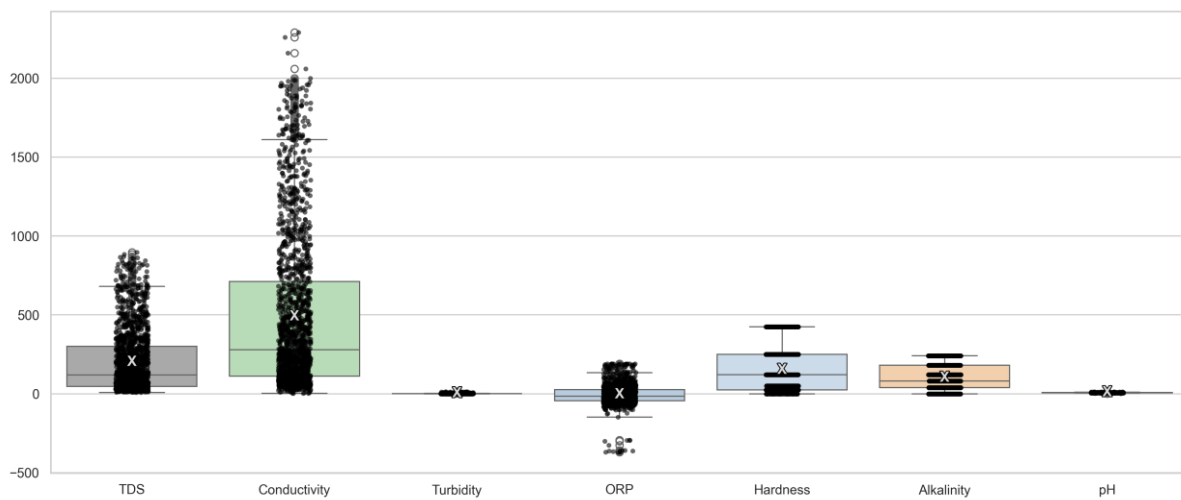

**Figure 7**. Distribution of physicochemical water quality parameters across all household samples, showing overall variations.

## 3.6 Cross Verification and Validation

Ensuring the reliability of non-expert contributions in community-based water monitoring requires structured validation and oversight. Several studies have demonstrated the potential of citizen-collected data when combined with rigorous study design and expert validation, an approach systematically adopted in the PWD initiative.

See et al. 2013, found that non-expert accuracy improves with expert validation, confidence scoring, and cross-comparison with expert-labelled data. Their study also highlighted that higher self-reported confidence correlates with greater accuracy, reinforcing the value of feedback mechanisms in crowdsourced monitoring. Ramesh et al. 2024, examined the reliability of water quality monitoring conducted by trained women in low-resource settings, demonstrating that non-experts can play a meaningful role in generating valid data when supported by well-structured and simplified tools. The findings resonate with the present study in two key respects. First, both studies underscore the importance of clarity and structure in survey design. Closed-ended, clearly phrased questions based on observable parameters consistently yielded higher agreement rates, supporting the use of simple, standardized formats in community-based surveys. Second, both studies reveal meaningful variation across question types: open-ended or interpretive questions, as well as those requiring dynamic judgement, tend to produce lower consistency. This indicates that beyond training alone, methodological simplification remains essential for ensuring reliability, particularly when scaling up student- or community-led data collection. These insights are further reinforced by USAID's pilot study in Tanzania, which similarly highlighted the promise of decentralized water monitoring while cautioning about the challenges of validation and standardization.[20]

Findings from the People's Water Data initiative support this, showing that *E. coli* detection rates in non-expert samples matched expert data in 96% of cases, demonstrating strong agreement. Details on additional parameters and full agreement rates across all survey questions are detailed in **Text S4** in the Supporting Information. Here, in **Figure 8**, we present a selection of representative questions to discuss agreement patterns. Additionally, adherence to GIS-tracked survey protocols increased significantly from 60% to 90% following the implementation of real-time tracking mechanisms and the inclusion of multi-angle photographs capturing both the street environment and household entrance. This improvement underscores the effectiveness of structured oversight and enhanced visual documentation in ensuring protocol compliance.

When comparing the responses collected by students and experts across key survey questions, clear patterns emerge regarding which types of questions lead to high agreement and where discrepancies occur. Questions based on simple and visible observations, such as the type of water filter, whether the water is stored in a container, the placement of the container, or the number of children under five, showed very high agreement rates, ranging from 93% to 100%. This suggests that students are generally reliable when it comes to recording clear, objective, and observable data in the field. In contrast, questions that required attention to actions or behaviors, such as how water was drawn from the container or whether the respondent's hand touched the water, had more moderate agreement levels, around 72% to 82%. These types of questions appear to require greater focus and real-time observation skills and are more prone to individual interpretation or missed details. The question with the lowest agreement was the name of the respondent, with only 48% alignment. This likely reflects issues with how students recorded the name, such as spelling, transcription errors, or the presence of different individuals during each visit. On the other hand, a personal numerical detail like the number of children was captured with high accuracy, reinforcing the idea that students manage structured or quantitative data more effectively. Interestingly, even questions related to risk perception and household health, which are more subjective in nature, had high agreement rates between 85% and 90%. This indicates that students were able to explain the questions clearly and document self-reported information in a consistent manner.

Overall, the findings suggest that students can be trusted to collect accurate and reliable data, especially for factual or observational questions. However, agreement levels tend to drop when dealing with open-ended questions, which introduce more room for variation in how responses are interpreted and recorded. In contrast, multiple-choice questions produce more consistent results and higher agreement rates. To further improve the reliability of student-led data collection, additional training should be provided with a focus on observational skills and accuracy in recording, alongside a revision of survey design to reduce ambiguity and ensure greater standardization in question formats.

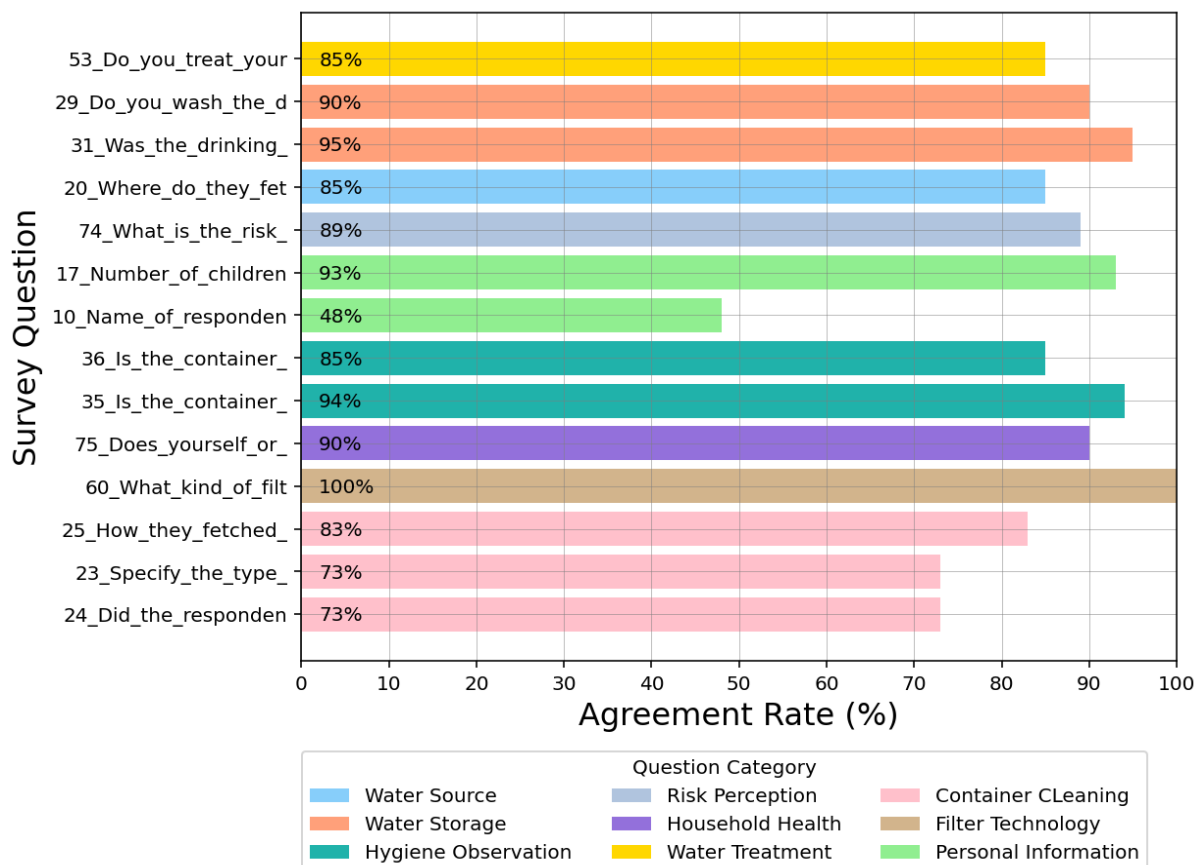

**Figure 8**. Agreement rates between student and expert responses across key household water survey questions, categorized by topic. The full questions are detailed in Table S7.

Water quality monitoring is essential for ensuring public health, yet significant gaps exist in many parts of the world, particularly in low-resource settings. Crowdsourced platform offers a potential solution by empowering local populations to contribute to data collection efforts.

One of the key challenges in drinking water quality assessment is the lack of systematic, large-scale monitoring, particularly in low-resource settings. While traditional models struggle with infrastructure and engagement, crowdsourcing, when supported by digital tools and expert verification, offers a promising path forward. The Contextual Framework for Crowdsourcing Water Quality Data outlines a structured approach for integrating community-driven environmental monitoring into water quality assessment, emphasizing public participation, technological accessibility, and validation mechanisms to make such systems effective and ensure data reliability.[21]

# 4. Conclusions

This study examined the opportunities and limitations of student-led, crowdsourced approaches to monitor drinking water quality through the People's Water Data (PWD) initiative. The findings demonstrate that engaging non-expert contributors, particularly students, offers a promising strategy for expanding the geographic scope and temporal frequency of water quality monitoring, particularly in low-resource or under-surveyed areas. Within two years, student monitors assessed over 9,000 household water sources across India, Uganda, and Israel, showcasing the feasibility of decentralized, community-driven monitoring at scale.

Several operational and methodological challenges were identified. These included spatial clustering of sampling points, inconsistent adherence to protocols, and context-specific barriers such as accessibility, safety, and respondent availability. These factors can compromise completeness, consistency, and representativeness. To address such issues, the initiative incorporated multiple quality assurance measures, most notably real-time GIS tracking, automated time-stamping, expert cross-validation, and structured feedback loops. These strategies were informed by prior research showing that non-expert accuracy improves significantly when supported by expert oversight and confidence scoring mechanisms.[10,15]

The study found that data reliability was closely linked to the type and structure of survey questions. Consistent with previous research,[5] closed-ended, clearly worded, and observable questions achieved higher agreement rates between student and expert assessments, often exceeding 90%. In contrast, open-ended or interpretive questions yielded more variable responses, underscoring the need for methodological simplification when scaling community-based monitoring. As demonstrated by the CrowdWater Game,[19] repeated assessments and consensus scoring can help identify anomalies and reinforce data quality in crowdsourced environments. These principles align with the PWD model, which similarly employed systematic comparisons and real-time validation mechanisms to improve non-expert contributions.

From a broader perspective, the PWD initiative addresses a critical global challenge: the persistent lack of systematic water quality monitoring in many parts of the world. Traditional monitoring systems often face limitations in infrastructure, financing, and local engagement especially in remote or underserved communities. Crowdsourcing, when embedded within a structured framework of public participation, technological accessibility, and validation mechanisms,[21] offers a scalable and sustainable alternative. The PWD model reflects this framework by combining student engagement, digital innovation, expert validation, and institutional partnerships.

Looking ahead, the continued development of flexible, modular methodologies, tailored to diverse geographic and socio-economic settings, is key to further scaling the PWD initiative. Integrating machine learning for predictive analysis and anomaly detection, as well as expanding real-time feedback loops for fieldworkers, may further enhance scientific rigor. Moreover, strategic collaborations with governments, NGOs, and academic institutions will be essential for embedding such approaches into formal water governance systems, thereby maximizing their long-term impact. The PWD initiative demonstrates how decentralized, people-powered monitoring can complement formal water governance systems and promote equitable, sustainable access to safe drinking water. We aim to scale up to one million mapped water sources through the participation of 100,000 trained students.

In summary, the PWD initiative demonstrates that when properly designed, supported, and validated, student-led and community-based monitoring can generate scientifically robust, policy-relevant data. It offers a blueprint for democratizing environmental monitoring and fostering inclusive, data-driven water governance in an era where local action and global sustainability are increasingly interconnected.

All data and results from this study are openly accessible on the People's Water Data platform, ensuring transparency, reproducibility, and global knowledge sharing.
https://www.peopleswaterdata.org/.

# 5. Funding Declaration

The initial pilot activities of this work were supported by the ASPER Clean Water Foundation (https://asperfoundation.com/asper-clean-water-fund) and the Olessia Kantor Fund (https://www.olessia-kantor.com). Following the success of the pilot, the formal creation and continued development of the People's Water Data initiative were made possible through the support of Petronet LNG Limited under its Corporate Social Responsibility (CSR) grant.

T. Pradeep thanks Petronet LNG Limited for supporting the broader vision of democratizing access to clean and safe water through academic community partnerships.

# 6. Acknowledgments

The initiative is supported by a robust network of academic institutions, including the primary academic partners: the Indian Institute of Technology (IIT) Madras, Tel Aviv University, and KMCH Research Foundation. The course titled "A Hybrid Course on Water Quality - An Approach to People's Water Data," combining classroom instruction in hybrid mode combined with practical field analysis was approved the Institutional Ethics Committee (IEC) of the Indian Institute of Technology Madras (Approval No. IEC/2024-02/PT/16, valid from 21 June 2024 to 20 June 2027). The course and related resources are accessible *via* NPTEL. We extend our gratitude to the 1,690 students who participated in this initiative in the past 3 semesters, which demonstrated the potential of student-led water quality monitoring. This initiative aligns with our vision of democratizing water education and represents a significant step toward enhancing quality, equity, and innovation in water management. The collaborating institutions across India are Dr NGP Arts and Science College, Fatima College, KPR Institute of Engineering and Technology, Madras Christian College, Mar Ivanois College, Mata Gujri Mahila Mahavidyalaya, SRKR Engineering College, Stella Maris College, and multiple government colleges nationwide.

The initiative also works closely with key NGO and public sector organizations, including NALA Foundation (Ethiopia), Africa Innovation Institute (Uganda), Development Alternatives (India), Engineering Without Borders (India chapter), and Jal Jeevan Mission (Assam, India). These partnerships represent a unified mission and an interdisciplinary effort to advance sustainability, inclusion, and human development through water literacy and community empowerment.

Data availability: The course materials, including lecture notes, lecture videos and student interactions across all the semesters are available via NPTEL (https://nptel.ac.in). The data collected are made available to public through the People's Water Data platform, https://www.peopleswaterdata.org developed by Sankar Sudhir as part of his MS thesis work, which will be published separately.

Guest lectures were delivered by experts from collaborating institutions, as listed in **Text S1** in the Supporting Information, including faculty from IIT Madras, Tel Aviv University, and the KMCH Research Foundation.

**Author Contributions**

S.K. initiated and led the methodological development of the People's Water Data (PWD) initiative, including the design, pilot testing, and refinement of core research tools. She co-developed the household survey tool with undergraduate students from Tel Aviv University, conducted the first field pilot, and co-led the subsequent pilot in Nallampatti village alongside G.V. As the course coordinator and jointly supervised Ph.D. student of H.M. and T.P., she served as the lead research and primary author of this manuscript, overseeing its structure, writing, and integration of field insights S.S. developed the digital platform for People's Water Data as part of his MS thesis work and, together with G.V., migrated the survey tool from Qualtrics to EpiCollect5, establishing the data architecture of the PWD system. T.L. participated in the early pilot phases, including the Nallampatti fieldwork, and played a key role in building and refining the Qualtrics survey tool, incorporating upgrades that shaped its current version. S. Seth trained the first cohort of student participants, contributed actively to the Nallampatti pilot, and supported heavy metal measurement and analysis. T.N. joined the Nallampatti field pilot and contributed to heavy metal testing activities. B.H. supported data analysis and field process optimization, contributing to the polygon-based randomization model and spatial design for survey allocation. R.D. coordinated the academic course at IIT Madras, and P.A. oversaw field operations in Coimbatore. G.V. served as a lecturer in the theoretical component, co-developed the EpiCollect5 survey and health modules, managed field logistics, and implemented the transition from Aquagenx to HiMedia biological test kits to improve cost efficiency while maintaining data quality. R.F. provided mentorship and guidance on statistical validation and randomization processes. A.M. coordinated fieldwork in Israel alongside S.K. and supported manuscript preparation and content refinement. H.M. supervised S.K., ensuring methodological consistency and scientific rigor. H.M. was closely involved throughout the development and implementation of the initiative. She participated directly in field activities during the early pilots, provided continuous scientific support, and offered hands-on supervision and feedback across all stages. T.P. conceptualized and led the People's Water Data initiative, including its course content, structure, coordination, visualization, and scale-up. He supervised the project overall, integrating its scientific, educational, and societal components.

11. Bouman, L., Spuhler, D., Bünzli, M.-A., Melad, A. III, Diop, L., Coelho, O., & Meierhofer, R. (2024). The water flow diagram. Frontiers in Water, 6, 1360515. https://doi.org/10.3389/frwa.2024.1360515.

12. Rodriquez, C., O'Connor, S. K., & Albers, E. (2025, Mar-Apr). American Pharmacists Association Foundation Incentive Grants: A 30-year descriptive review. Journal of the American Pharmacists Association, 65(2), Article 102323. https://doi.org/10.1016/j.japh.2025.102323.

13. Bis, I. (2012). Bureau of Indian Standards. New Delhi, 6, 2-3.

14. Rice, E. W., Baird, R. B., Eaton, A. D., & Clesceri, L. S., et al. (2012). Standard methods for the examination of water and wastewater (22nd ed.). American Public Health Association (APHA).

15. Clasen, T. F., & Bastable, A. (2003). Faecal contamination of drinking water during collection and household storage: the need to extend protection to the point of use. Journal of water and health, 1(3), 109-115.

16. Mintz, E. D., Reiff, F. M., & Tauxe, R. V. (1995). Safe water treatment and storage in the home: a practical new strategy to prevent waterborne disease. Jama, 273(12), 948-953.

17. Lindskog, R., & Lindskog, P. (1988). Bacteriological contamination of water in rural areas: an intervention study from Malawi. Tropical Medicine & International Health, 91(1), 1-7.

18. Ajith, V., Fishman, R., Yosef, E., Edris, S., Ramesh, R., Suresh, R. A., Pras, A., Rahim, V., Rajendran, S., & Yanko, M. (2023). An integrated methodology for assessment of drinking-water quality in low-income settings. Environmental Development, 46, 100862.

19. Strobl, B., Etter, S., van Meerveld, I., & Seibert, J. (2019). The CrowdWater game: A playful way to improve the accuracy of crowdsourced water level class data. *PLOS ONE, 14*(9), e0222579. https://doi.org/10.1371/journal.pone.0222579.

20. USAID 2016, Pilot study Tanzania.

21. Mistry, J., Li, J., Yoshikawa, H., Tseng, V., Tirrell, J., Kiang, L., Mistry, R., & Wang, Y. (2016). An integrated conceptual framework for the development of Asian American children and youth. Child Development, 87(5), 1366–1385. https://doi.org/10.1111/cdev.12577.
29